\documentclass[11pt,a4paper]{article}
\usepackage{jheppub}
\usepackage{amsmath}
\usepackage[most]{tcolorbox}
\usepackage{dsfont}
\usepackage{ulem}
\usepackage{natbib}
\usepackage{xcolor}
\usepackage[hang,flushmargin]{footmisc}
\usepackage{tikz-cd}
\usepackage{enumitem}
\usepackage{amsfonts,amssymb, amscd,amsmath,latexsym,amsbsy,bm}
\usepackage{stmaryrd}
\usepackage{todonotes}
\usepackage{stackrel}
\usepackage{amsthm}

\usepackage{graphicx}

\let\zeta\xi

\title{Variational symmetries of Lagrangian systems with higher-order derivatives}

\author{Ege Coban$^a$, Ilmar Gahramanov$^{a,b,c}$, and Dilara Kosva$^{a}$}
\affiliation{
	$^a$ {Department of Physics, Bogazici University, 34342 Bebek, Istanbul, Turkey}\\[-0.5cm]
	
	$^{b}$ Institute of Radiation Problems, Azerbaijan National Academy of Sciences,  B.Vahabzade St. 9, AZ1143, Baku, Azerbaijan\\[-0.5cm]
	
	$^{c}$ Department of Mathematics, Khazar University,  Mehseti St. 41, AZ1096, Baku, Azerbaijan \\[-0.5cm]
}

{\footnotesize \emailAdd{ege.coban@boun.edu.tr}
\emailAdd{ilmar.gahramanov@boun.edu.tr}
\emailAdd{dilara.kosva@boun.edu.tr}
}

	\abstract{
		
		We discuss an elementary derivation of variational symmetries and corresponding integrals of motion for the Lagrangian systems depending on acceleration.  Providing several examples, we make the manuscript accessible to a wide range of readers with interest in higher-order Lagrangians and symmetries. The discussed technique is also applicable to the Lagrangian systems with higher-order derivatives.}

\begin{document}
	\maketitle
	\flushbottom

	\section{Introduction}
	
	There are many reasons why physicists and mathematicians are interested in studying Lagrangian systems with higher-order derivatives. The history of such theories dates back to the study of instabilities due to Ostrogradsky \cite{ostrogradsky1850memoires}. Since then, theories with higher-order derivatives have been studied by many authors from various view points\footnote{It is impossible to cite all relevant papers, we mention here some more modern works for the interested reader \cite{Woodard:2015zca,de1995symmetries,ccaugatay2018reductions, cruz2016hamiltonian}.}. 
	
	In \cite{torres2013variational} (see also \cite{del2017variational,arutyunov2019liouville}) authors presented an elementary derivation of the variational symmetries and the integrals of motion associated with them. In this paper, we will consider an extension of this idea to the Lagrangian systems with higher-order derivatives, especially to the case when the Lagrangian depends on generalized acceleration. Essentially, we give a derivation of the equation that determines variational symmetries and then obtain integrals of motion for some systems. To our knowledge, this technique is the simplest way to obtain integrals of motion for the Lagrangian systems with higher-order derivatives. It should be stressed that all computations presented here rely heavily on group-theoretical studies of differential equations \cite{ibragimov1995crc,ibragimov1992group,bluman2008symmetry,olver2000applications}.

	The paper is organized as follows. The variational symmetries and the relevant techniques are discussed in Sections II and III. Then we present some examples in Section IV, where we mainly focus on the Lagrangians with second-order derivatives. In Section V, we shortly discuss higher orders.

	\section{Variational Symmetries for Second-Order Lagrangians}

	The Euler-Lagrange equation for the Lagrangian depending on second derivatives of generalized coordinates 
	$L=L(\mathbf{x},\dot{\mathbf{x}},\ddot{\mathbf{x}};t)$
	has the following form,
	\begin{equation}
		\frac{\partial L}{\partial x_i}-\frac{\mathrm{d}}{\mathrm{d}t}\frac{\partial L}{\partial \dot{x}_i}+\frac{\mathrm{d}^2}{\mathrm{d}t^2}\frac{\partial L}{\partial \ddot{x}_i}=0\; .
	\end{equation}
	This is a fourth-order ordinary differential equation. We are going to construct variational symmetries for this Lagrangian system by solving some corresponding differential equations.
	
	Consider an action $S=\int L(\mathbf{x},\dot{\mathbf{x}},\ddot{\mathbf{x}};t) \; \mathrm{d}t$ which is invariant under the invertible $s$-parametric transformations $x'_i=x'_i(\mathbf{x},t,s)$ and $t'=t'(\mathbf{x},t,s)$. These transformations form a one-parameter group (with the parameter $s$). The corresponding infinitesimal transformations are
	\begin{align}\label{x'}
		x'_i = x_i +\eta_i(\mathbf{x},t) s \hspace{.5cm};\hspace{.5cm}i=1,2,...,n\\\label{t'}
		t'=t+\zeta(\mathbf{x},t) s \hspace{.5cm};\hspace{.5cm}i=1,2,...,n\;,
	\end{align}
	where the functions $\eta_i$ and $\zeta$ are given by
	\begin{equation}
		\eta_i(\mathbf{x};t)=\frac{\partial x_i'(\mathbf{x},t,s)}{\partial s}\Bigr\rvert_{s=0}\hspace{.5cm};\hspace{.5cm}i=1,2,...,n
	\end{equation}
	\begin{equation}
		\zeta(\mathbf{x};t)=\frac{\partial t'(\mathbf{x},t,s)}{\partial s}\Bigr\rvert_{s=0}\hspace{.5cm};\hspace{.5cm}i=1,2,...,n\;.
	\end{equation}
	The invariance of the action under these transformations (in order to obtain the same equations of motion) implies that
	\begin{equation} \label{varsymLag}
		L'(\mathbf{x}',\frac{\mathrm{d}\mathbf{x}'}{\mathrm{d}t'},\frac{\mathrm{d}^2\mathbf{x}'}{\mathrm{d}t'^2},t')\frac{\mathrm{d}t'}{\mathrm{d}t}=L(\mathbf{x},\dot{\mathbf{x}},\ddot{\mathbf{x}},t)+\frac{\mathrm{d}}{\mathrm{d}t}F(\mathbf{x},\dot{\mathbf{x}},t,s)\;,
	\end{equation}
	where we added the total derivative term to the Lagrangian. If the Lagrangian transforms under the transformations (\ref{x'})-(\ref{t'}) as in expression (\ref{varsymLag}) then the corresponding symmetry is called a variational symmetry of Lagrangian. By differentiating the last expression  with respect to the parameter $s$ and by setting $s=0$ one obtains
	\begin{equation}\label{eq1}
		\sum_{i=1}^n\left(\frac{\partial L}{\partial x_i}\eta_i+\frac{\partial L}{\partial \dot{x}_i}\Bigr[\frac{\mathrm{d}\eta_i}{\mathrm{d}t}-\dot{x}_i\frac{\mathrm{d}\zeta}{\mathrm{d}t}\Bigr]+\frac{\partial L}{\partial \ddot{x}_i}\Bigr[\frac{\mathrm{d}^2\eta_i}{\mathrm{d}t^2}-2\ddot{x}_i\frac{\mathrm{d}\zeta}{\mathrm{d}t}-\dot{x}_i\frac{\mathrm{d}^2\zeta}{\mathrm{d}t^2}\Bigr]\right)+\frac{\partial L}{\partial t}\zeta+L\frac{\mathrm{d}\zeta}{\mathrm{d}t}=\frac{\mathrm{d}}{\mathrm{d}t} G(\mathbf{x},\dot{\mathbf{x}};t)\;.
	\end{equation}
	For the terms inside the square brackets we used the following expressions
	\begin{eqnarray}
		\frac{\partial \dot{x}_{i}'}{\partial s}\Bigr\rvert_{s=0}&=&\frac{\mathrm{d} \eta_{i}}{\mathrm{d} t}-\dot{x}_{i} \frac{\mathrm{d} \zeta}{\mathrm{d} t}\\
		\frac{\partial\ddot{x}_i'}{\partial s}\Bigr\rvert_{s=0} &=& \frac{\mathrm{d}^2 \eta_i}{\mathrm{d}t^2}-2\ddot{x}_i\frac{\mathrm{d} \zeta}{\mathrm{d} t} - \dot{x}_i\frac{\mathrm{d}^2 \zeta}{\mathrm{d}t^2} \; .
	\end{eqnarray}
	and the gauge function $G(\mathbf{x},\dot{\mathbf{x}};t)$ is defined as 
	\begin{equation}
		G(\mathbf{x}, \dot{\mathbf{x}}; t) :=\frac{\partial }{\partial s} F(\mathbf{x},\dot{\mathbf{x}},t,s) \Bigr\rvert_{s=0}\;.
	\end{equation}
	Now by using the Euler-Lagrange equation in $\eqref{eq1}$, one ends up with the following expression
	\begin{equation}\label{Q}
		\frac{\mathrm{d}}{\mathrm{d}t}  \left[\zeta L + \sum_{i=1}^{n}\left((\eta_i - \dot{x}_i\zeta) \frac{\partial L}{\partial \dot{x}_i} + \frac{\mathrm{d}(\eta_i - \dot{x}_i\zeta) }{\mathrm{d}t} \frac{\partial L}{\partial \ddot{x}_i} - (\eta_i - \dot{x}_i\zeta) \frac{\mathrm{d}}{\mathrm{d}t}\left(\frac{\partial L}{\partial \ddot{x}_i}\right)\right)-G\right] = 0 \;.
	\end{equation}
	The expression in the bracket is conserved, i.e. it is integral of motion under the time evolution \cite{torres2013variational,deriglazov2016classical}. This expression is an analog of the Rund-Trautman identity\footnote{In the literature, sometimes this expression is called Noether-Bassel-Hagen identity, see e.g. \cite{leone2018wonderfulness}.} \cite{trautman1967noether,neuenschwander2017emmy,gourieux2021noether}.
	
	Now let us compute variational symmetries for some Lagrangian systems with the second-order derivatives. Using the expression (\ref{eq1}) one can easily derive integrals of motion for relatively simple systems.

	\subsection{Spinning particle}
	
	We begin with the most familiar example of the second order Lagrangian systems, namely with the Lagrangian of the classical spinning particle (for simplicity we put $m=\omega=1$)
	\begin{equation}
		L=\frac{1}{2}(\ddot{x}^2-\dot{x}^2) \;.
	\end{equation}
	The Euler-Lagrange equation for this system is the following fourth-order differential equation  which describes a particle rotating around its translating center
	\begin{equation}
		x^{(4)}+\ddot{x}=0    \;.
	\end{equation}
	This system was discussed in many works, the integrals of motion and related symmetries can be found, e.g. in \cite{cracsmuareanu2000noetherian}.
	By inserting the Lagrangian of the spinning particle to the expression (\ref{eq1}) yields
	\begin{equation}
		\begin{split}
			\ddot{x}^2\Bigr[\frac{\partial \eta}{\partial x}-\frac{5}{2}\dot{x}\frac{\partial \zeta}{\partial x}-\frac{3}{2}\frac{\partial \zeta}{\partial t}\Bigr]+\ddot{x}\Bigr[\dot{x}^2\frac{\partial^2 \eta}{\partial x^2}+2\dot{x}\frac{\partial^2 \eta}{\partial t\partial x}+\frac{\partial^2 \eta}{\partial t^2}-\dot{x}^3\frac{\partial^2\zeta}{\partial x^2}-2\dot{x}^2\frac{\partial^2\zeta}{\partial t\partial x}\\-\dot{x}\frac{\partial^2\zeta}{\partial t^2}\Bigr]+\dot{x}^3\frac{1}{2}\frac{\partial \zeta}{\partial x}+\dot{x}^2\Bigr[-\frac{\partial \eta}{\partial x}+\frac{1}{2}\frac{\partial \zeta}{\partial t}\Bigr]-\dot{x}\frac{\partial \eta}{\partial t}=\frac{\partial G}{\partial\dot{x}}\ddot{x}+\frac{\partial G}{\partial x}\dot{x}+\frac{\partial G}{\partial t}\;.
		\end{split}
	\end{equation}
	Since the gauge function $G(x,\dot{x};t)$ is not a function of $\ddot{x}$ coefficients of $\ddot{x}^2$ and $\ddot{x}$ terms should vanish. As a result we obtain the following differential equations
	\begin{equation}\label{firsteq}
		\frac{\partial \eta}{\partial x}-\frac{5}{2}\dot{x}\frac{\partial \zeta}{\partial x}-\frac{3}{2}\frac{\partial \zeta}{\partial t}=0
	\end{equation}
	\begin{equation}\label{secondeq}
		\dot{x}^2\frac{\partial^2 \eta}{\partial x^2}+2\dot{x}\frac{\partial^2 \eta}{\partial t\partial x}+\frac{\partial^2 \eta}{\partial t^2}-\dot{x}^3\frac{\partial^2\zeta}{\partial x^2}-2\dot{x}^2\frac{\partial^2\zeta}{\partial t\partial x}-\dot{x}\frac{\partial^2\zeta}{\partial t^2}=\frac{\partial G}{\partial \dot{x}}
	\end{equation}
	\begin{equation}\label{thirdeq}
		\dot{x}^3\frac{1}{2}\frac{\partial \zeta}{\partial x}+\dot{x}^2\Bigr[-\frac{\partial \eta}{\partial x}+\frac{1}{2}\frac{\partial \zeta}{\partial t}\Bigr]-\dot{x}\frac{\partial \eta}{\partial t}=\frac{\partial G}{\partial x}\dot{x}+\frac{\partial G}{\partial t}\;.
	\end{equation}
	In some sense, these equations have the same complexity as the Euler-Lagrange equations, here the idea is to find symmetries and corresponding integrals of motion. The equation (\ref{firsteq}) implies that $\zeta(x;t)$ is not a function of $x$ and moreover $\eta(x;t)$ can be found as
	\begin{equation}
		\eta=\frac{3}{2}x\frac{\mathrm{d}\zeta}{\mathrm{d}t}+A(t)\;,
	\end{equation}
	where $A(t)$ is a function of time only. If we insert the expression for $\eta$ in the expressions (\ref{secondeq})-(\ref{thirdeq}), we obtain the following equations for the partial differentials of the gauge function $G(x,\dot{x};t)$ 
	\begin{equation}
		-2\dot{x}\frac{\mathrm{d}\zeta}{\mathrm{d}t}-\frac{3}{2}x\frac{\mathrm{d}^2\zeta}{\mathrm{d}t^2}-\frac{\mathrm{d}A}{\mathrm{d}t}-\frac{7}{2}\dot{x}\frac{\mathrm{d}^3\zeta}{\mathrm{d}t^3}-\frac{3}{2}x\frac{\mathrm{d}^4\zeta}{\mathrm{d}t^4}-\frac{\mathrm{d}^3A}{\mathrm{d}t^3}=\frac{\partial G}{\partial x}
	\end{equation}
	\begin{equation}
		2\dot{x}\frac{\mathrm{d}^2\zeta}{\mathrm{d}t^2}+\frac{3}{2}x\frac{\mathrm{d}^3\zeta}{\mathrm{d}t^3}+\frac{\mathrm{d}^2A}{\mathrm{d}t^2}=\frac{\partial G}{\partial\dot{x}}
	\end{equation}
	\begin{equation}
		\dot{x}^2\frac{\mathrm{d}\zeta}{\mathrm{d}t}+\frac{7}{2}\dot{x}^2\frac{\mathrm{d}^3\zeta}{\mathrm{d}t^3}+\frac{3}{2}x\dot{x}\frac{\mathrm{d}^4\zeta}{\mathrm{d}t^4}+\dot{x}\frac{\mathrm{d}^3A}{\mathrm{d}t^3}=\frac{\partial G}{\partial t}\;.
	\end{equation}
	By comparing them one obtains the following four ordinary differential equations
	\begin{equation}
		2\frac{\mathrm{d}\zeta}{\mathrm{d}t}+5\frac{\mathrm{d}^3\zeta}{\mathrm{d}t^3}=0
	\end{equation}
	\begin{equation}
		2\frac{\mathrm{d}^2\zeta}{\mathrm{d}t^2}+5\frac{\mathrm{d}^4\zeta}{\mathrm{d}t^4}=0
	\end{equation}
	\begin{equation}
		\frac{\mathrm{d}^3\zeta}{\mathrm{d}t^3}+\frac{\mathrm{d}^5\zeta}{\mathrm{d}t^5}=0
	\end{equation}
	\begin{equation}
		\frac{\mathrm{d}^2A}{\mathrm{d}t^2}+\frac{\mathrm{d}^4A}{\mathrm{d}t^4}=0\;.
	\end{equation}
	One can straightforwardly solve these differential equations, and find that
	\begin{equation}
		\zeta=c_1\hspace{.5cm};\hspace{.5cm}A=c_2\sin{t}+c_3\cos{t}+c_4t+c_5\;.
	\end{equation}
	Therefore $\eta(x;t)$ has the form 
	\begin{equation}
		\eta=c_2\sin{t}+c_3\cos{t}+c_4t+c_5\;,
	\end{equation}
	where $c_2$, $c_3$, $c_4$ and $c_5$ are arbitrary constants. It is easy to compute the gauge function $G(x,\dot{x};t)$ 
	\begin{equation}
		G=-c_2\dot{x}\sin{t}-c_3\dot{x}\cos{t}-c_4x\;.
	\end{equation}
	Now by inserting $\eta$, $\zeta$ and $G$ into the (\ref{Q}) one ends up with the following expression
	\begin{equation}
		\begin{split}
			\frac{\mathrm{d}}{\mathrm{d}t}\Bigr[c_1(-\frac{\ddot{x}^2}{2}+\frac{\dot{x}^2}{2}+\dot{x}\dddot{x})+c_2(\ddot{x}\cos{t}-\dddot{x}\sin{t})+c_3(-\dddot{x}\cos{t}-\ddot{x}\sin{t})\\+c_4(-\dot{x}t-\dddot{x}t+\ddot{x}+x)+c_5(-\dot{x}-\dddot{x})\Bigr]=0\;.
		\end{split}
	\end{equation}
	Since $c_1$, $c_2$, $c_3$, $c_4$ and $c_5$ are arbitrary constants, terms within the parenthesis are integrals of motion with the following symmetry generators
	\begin{alignat}{3}
		\Gamma_1&=\frac{\partial}{\partial t}\hspace{.5cm}&&;\hspace{.5cm}I_1&&=\ddot{x}^2-2\dot{x}\dddot{x}-\dot{x}^2 \\
		\Gamma_2&=\sin{t}\frac{\partial}{\partial x}\hspace{.5cm}&&;\hspace{.5cm}I_2&&=\ddot{x}\cos{t}-\dddot{x}\sin{t} \\
		\Gamma_3&=\cos{t}\frac{\partial}{\partial x}\hspace{.5cm}&&;\hspace{.5cm}I_3&&=\ddot{x}\sin{t}+\dddot{x}\cos{t} \\
		\Gamma_4&=t\frac{\partial}{\partial x}\hspace{.5cm}&&;\hspace{.5cm}I_4&&=x+\ddot{x}-t(\dot{x}+\dddot{x}) \\
		\Gamma_5&=\frac{\partial}{\partial x}\hspace{.5cm}&&;\hspace{.5cm}I_5&&=\dot{x}+\dddot{x}\;.
	\end{alignat}
	Note that in the two-dimensional case, i.e. if the Lagrangian of the spinning particle has a form
	\begin{equation}
		L=\frac{1}{2}(\ddot{x}^2-\dot{x}^2+\ddot{y}^2-\dot{y}^2)\;,
	\end{equation}
	we will get one more integral of motion (the third component of the angular momentum)
	\begin{equation}
		\Gamma_6=x\frac{\partial}{\partial y}-y\frac{\partial}{\partial x}\hspace{.5cm};\hspace{.5cm}I_6=y(\dot{x}+\dddot{x})-x(\dot{y}+\dddot{y})+\ddot{y}\dot{x}-\ddot{x}\dot{y}\;.
	\end{equation}
	
	As one can see, we constructed the infinitesimal transformations by solving corresponding  differential equations\footnote{Of course, if one knows the symmetries in advance, then it is easier to use the Noether theorem to obtain the integrals of motion.}. 
	
	\section{Cyclic coordinates}
	
	If we have a one-parameter symmetry group of the action, then we can discard one of the coordinates and consider the parameter $s$ as a new coordinate. Then the coordinate $s$ turns out to be cyclic (ignorable) and the corresponding momentum is conserved.  Actually, this is an idea of Noether's theorem.
	
	We would like to show that for $L(\mathbf{x},\mathbf{\Dot{x}},\mathbf{\Ddot{x}};t)$, there exists a coordinate system $({\mathbf{x}}', t')$ such that for a fixed $k$ the momentum conjugate corresponding to  ${x_k}'$ is conserved.
	\begin{equation}
		\frac{\mathrm{d}p'_k}{\mathrm{d}t'} = 0 \; .
	\end{equation}
	Under coordinate transformations $x_i= x_i({\mathbf{x}}',t')$ and $t = t({\mathbf{x}}',t')$, Lagrangian transforms as;
	\begin{equation}\label{lag transform}
		L(\mathbf{x},\mathbf{\Dot{x}},\mathbf{\Ddot{x}};t) \frac{\mathrm{d}t}{\mathrm{d}t'} = L'(\mathbf{x}',\mathbf{\Dot{x}}',\mathbf{\Ddot{x}}';t)
	\end{equation}
	\begin{equation}\label{p=0}
		\frac{\partial L'}{\partial {x_k}'} = L \frac{\partial}{\partial x_{k}^{\prime}}\left(\frac{\mathrm{d} t}{\mathrm{d} t^{\prime}}\right)+\frac{\mathrm{d} t}{\mathrm{d} t^{\prime}} \left(\sum_{i=1}^{n}\left(\frac{\partial L}{\partial x_{i}} \frac{\partial x_{i}}{\partial x_{k}^{\prime}}+\frac{\partial L}{\partial \dot{x}_{i}} \frac{\partial \dot{x}_{i}}{\partial x_{k}^{\prime}}+ \frac{\partial L}{\partial \ddot{x_i}}\frac{\partial\ddot{x_i}}{\partial {x_k}'}\right)+\frac{\partial L}{\partial t} \frac{\partial t}{\partial x_{k}^{\prime}}  \right)=0 \; .
	\end{equation}
	In order to simplify this expression, we make the following definitions and follow the same method used in Section II,
	\begin{equation}
		\zeta:=\frac{\partial t}{\partial x_{k}'} \; \;     \eta_{i} := \frac{\partial x_i}{\partial x_k'}
	\end{equation}
	\begin{equation}\label{asil}
		L \frac{\mathrm{d} \zeta}{\mathrm{d} t}+\sum_{i=1}^{n}\left[\frac{\partial L}{\partial x_{i}} \eta_{i}+\frac{\partial L}{\partial \dot{x}_{i}}\left(\frac{\mathrm{d} \eta_{i}}{\mathrm{d} t}-\dot{x}_{i} \frac{\mathrm{d} \zeta}{\mathrm{d} t}\right)+ \frac{\partial L}{\partial \ddot{x_i}}\left(\frac{\mathrm{d}^2 \eta_i}{\mathrm{d}t^2}-2\ddot{x_i}\frac{\mathrm{d} \zeta}{\mathrm{d} t} - \dot{x_i}\frac{\mathrm{d}^2 \zeta}{\mathrm{d}t^2}\right)\right]+\frac{\partial L}{\partial t} \zeta=0  \; ,
	\end{equation}
	Then the corresponding momentum takes the following form
	\begin{equation}
		p'_k = \zeta L + \sum_{i=1}^{n}(\eta_i - \dot{x}_i\zeta) \frac{\partial L}{\partial \dot{x}_i} + \frac{\mathrm{d}(\eta_i - \dot{x}_i\zeta) }{\mathrm{d}t} \frac{\partial L}{\partial \ddot{x}_i} - (\eta_i - \dot{x}_i\zeta) \frac{\mathrm{d}}{\mathrm{d}t}\left(\frac{\partial L}{\partial \ddot{x}_i}\right) \; .
	\end{equation}
	One can also consider the case where $x'_k$ is not necessarily cyclic. If,
	\begin{equation}\label{G}
		\frac{\partial L}{\partial x_k} = \frac{\mathrm{d}G}{\mathrm{d}t} \; ,
	\end{equation}
	Then from the Euler-Lagrange equations, 
	\begin{equation}
		\frac{\mathrm{d}(p_k -G)}{\mathrm{d}t} = 0  \;.
	\end{equation}
	In the extended configuration space the expression \eqref{G} corresponds to
	\begin{equation}
		L \frac{\mathrm{d} \zeta}{\mathrm{d} t}+\sum_{i=1}^{n}\left[\frac{\partial L}{\partial x_{i}} \eta_{i}+\frac{\partial L}{\partial \dot{x}_{i}}\left(\frac{\mathrm{d} \eta_{i}}{\mathrm{d} t}-\dot{x}_{i} \frac{\mathrm{d} \zeta}{\mathrm{d} t}\right)+ \frac{\partial L}{\partial \ddot{x_i}}\left(\frac{\mathrm{d}^2 \eta_i}{\mathrm{d}t^2}-2\ddot{x_i}\frac{\mathrm{d} \zeta}{\mathrm{d} t} - \dot{x}\frac{\mathrm{d}^2 \zeta}{\mathrm{d}t^2}\right)\right]+\frac{\partial L}{\partial t} \zeta= \frac{\mathrm{d}G}{\mathrm{d}t} \; .
	\end{equation}
	Hence the conserved quantity takes the form given in the expression \eqref{Q}. 
	
	It is still possible to write an equivalent Lagrangian $\Tilde{L}'$ where $x_k'$ is a cyclic coordinate. For that reason we define a function $F(\mathbf{x}', \dot{\mathbf{x}}'; t')$ as follows
	\begin{equation}\label{f}
		G = - \frac{\partial F}{\partial x'_k} \; .
	\end{equation}
	Now, it is easy to see that defining a new Lagrangian as $\tilde{L'}= L
	'+\dfrac{\mathrm{d} F}{\mathrm{d} t'}$ will result in an equivalent Lagrangian with a cyclic coordinate $x'_k$ since
	\begin{equation}\label{tilde L}
		\frac{\partial}{\partial x_{k}^{\prime}}\left(L^{\prime}+\frac{\mathrm{d} F}{\mathrm{d} t^{\prime}}\right)=0 \;.
	\end{equation}

	\section{Examples}
	
	\subsection{Higher-derivative harmonic oscillator}

	As an elementary example, we will consider the following second-order Lagrangian
	\begin{equation}\label{Lag1}
		L = \frac{1}{2}({\ddot{x}^2- x^2}) \; .
	\end{equation} 
	This system was studied as a toy model in \cite{Smilga:2005gb,Boulanger:2018tue}. We need to find variational symmetries satisfying the condition \eqref{asil}. Substituting the above Lagrangian into the equation \eqref{asil}, we obtain
	\begin{equation}
		-x\eta + \ddot{x}\left(\frac{\mathrm{d}^2 \eta_i}{\mathrm{d}t^2}-2\ddot{q_i}\frac{\mathrm{d} \zeta}{\mathrm{d} t} - \dot{q}\frac{\mathrm{d}^2 \zeta}{\mathrm{d}t^2}\right) + \frac{1}{2}\left(\ddot{x}^2- x^2 \right) \frac{\mathrm{d}\zeta}{\mathrm{d}t} = \frac{\mathrm{d}G}{\mathrm{d}t} \; , 
	\end{equation}
	where the gauge function $G(x,\dot{x};t)$ is a function of $t, x$ and $\dot{x}$. Following the same method that we have explained in our previous example, we can express our functions as follows
	\begin{eqnarray}
		\zeta &=& c_1 \\
		\eta &=& c_2 e^{t} + c_3 e^{-t} + c_4\cos{t} + c_5 \sin{t}\\
		G &=& c_2e^t(\dot{x}-x)+c_3e^{-t}(\dot{x}+x)+c_4(-\dot{x}\cos{t}-x\sin{t})+c_5(-\dot{x}\sin{t}+x\cos{t}) \; .
	\end{eqnarray}
	Corresponding symmetry generators and integrals of motion are
	\begin{alignat}{3}
		\Gamma_1&=\frac{\partial}{\partial t}\hspace{.5cm}&&;\hspace{.5cm}I_1&&=\dot{x}\dddot{x}-\frac{x^2}{2}-\frac{\ddot{x}^2}{2} \\
		\Gamma_2&=e^t\frac{\partial}{\partial x}\hspace{.5cm}&&;\hspace{.5cm}I_2&&=(x-\dot{x}+\ddot{x}-\dddot{x})e^t \\
		\Gamma_3&=e^{-t}\frac{\partial}{\partial x}\hspace{.5cm}&&;\hspace{.5cm}I_3&&=(x+\dot{x}+\ddot{x}+\dddot{x})e^{-t} \\
		\Gamma_4&=\cos{t}\frac{\partial}{\partial x}\hspace{.5cm}&&;\hspace{.5cm}I_4&&=(\dot{x}-\dddot{x})\cos{t}+(x-\ddot{x})\sin{t} \\
		\Gamma_5&=\sin{t}\frac{\partial}{\partial x}\hspace{.5cm}&&;\hspace{.5cm}I_5&&=(\dot{x}-\dddot{x})\sin{t}+(-x+\ddot{x})\cos{t}  \; .
	\end{alignat}
	In order to simplify the problem of finding a Lagrangian with a cyclic coordinate, we will consider the case where $c_i$'s are all equal to zero except $c_2$ which equals to $1$. Then
	\begin{equation}
		\zeta = 0 \; , \; \eta = e^{t} \; , \; G = \dot{x}e^{t} - x e^{t} \; .
	\end{equation}
	The coordinate transformations satisfying these equations are
	\begin{equation}
		t = t'  \; , \; x = x'e^{t'} \; .
	\end{equation}
	Under these transformations, the Lagrangian (\ref{Lag1}) transforms as
	\begin{equation}
		L' = \frac{1}{2}\left((\ddot{x}'e^{t'} + 2\dot{x}'e^{t'} + x'e^{t'})^2 - {x'}^2 e^{2t'} \right) \; .
	\end{equation}
	The Lagrangian $\tilde{L}'$ where $x'$ is a cyclic coordinate is given by
	\begin{equation}
		\tilde{L}' = L' + \frac{\mathrm{d} F}{\mathrm{d} t^{\prime}} \;,
	\end{equation}
	where $F(x, \dot{x}, t)$ is defined as in (\ref{f}). After expressing $G(x,\dot{x},t)$ and $F(x, \dot{x}, t)$ in terms of primed coordinates, we simply follow the definition and $\tilde{L}'$ takes the following form,
	
	\begin{equation}
		\tilde{L}^\prime = \frac{e^{2t^{\prime}}}{2}\left(\ddot{x}^{\prime 2} + 2 \dot{x}^{\prime 2} + 4 \ddot{x}^\prime\dot{x}^\prime \right) \;,
	\end{equation}
	where $x$ is clearly cyclic. The momentum conjugate corresponding to the coordinate $x$ is
	\begin{equation}
		p' = -e^{2t'}\left(\dddot{x\hspace{0pt}}' +  2 \Ddot{x}' + 2 \Dot{x}' \right ) \; .
	\end{equation}
	Returning to our initial coordinate system, the conserved quantity corresponds to, 
	\begin{equation}
		I = (x-\dot{x}+\ddot{x}-\dddot{x})e^t \; ,
	\end{equation}
	which is also given by \eqref{Q}.

	\subsection{An illustrative example}
	Now, we would like to move on to a slightly more complicated case. As a next example, consider a system with the following Lagrangian
	\begin{equation}\label{mat}
		L=\dot{x}^4+3x^2\ddot{x}^2\;.
	\end{equation}
	The Euler-Lagrange equation for this system is the following ordinary differential equation
	\begin{equation}
		3\ddot{x}^2+xx^{(4)}+4\dot{x}\dddot{x}=0\;.
	\end{equation}
	By applying (\ref{eq1}) we obtain the following partial differential equation
	\begin{equation}
		x\eta+x^2\frac{\partial \eta}{\partial x}-\frac{3}{2}x^2\frac{\mathrm{d}\zeta}{\mathrm{d}t}=0\;,
	\end{equation}
	from which one can easily find that $\zeta=\zeta(t)$. Then
	\begin{equation}
		\frac{\partial (x\eta)}{\partial x}=\frac{3}{2}x\frac{\mathrm{d}\zeta}{\mathrm{d}t}\;.
	\end{equation}
	Therefore $\eta(x;t)$ has the form $\eta=\dfrac{3}{4}x\dfrac{\mathrm{d}\zeta}{\mathrm{d}t}+\dfrac{A(t)}{x}$. 
	
	The partial differential equations involving derivatives of the gauge functions are
	\begin{equation}
		3x^2\dot{x}\frac{\mathrm{d}^2\zeta}{\mathrm{d}t^2}+\frac{9}{2}x^3\frac{\mathrm{d}^3\zeta}{\mathrm{d}t^3}+6x\frac{\mathrm{d}^2A}{\mathrm{d}t^2}-12\dot{x}\frac{\mathrm{d}A}{\mathrm{d}t}+12\frac{\dot{x}^2}{x}A=\frac{\partial G}{\partial \dot{x}}
	\end{equation}
	\begin{equation}
		3x\dot{x}^3\frac{\mathrm{d}^2\zeta}{\mathrm{d}t^2}+4\frac{\dot{x}^3}{x^2}\frac{\mathrm{d}A}{\mathrm{d}t}-4\frac{\dot{x}^4}{x^2}A=\frac{\partial G}{\partial x}\dot{x}+\frac{\partial G}{\partial t}\;.
	\end{equation}
	By solving these equations one finds $\zeta$ and $A$ as
	\begin{equation}
		\zeta=c_1t^2+c_2t+c_3\hspace{.5cm};\hspace{.5cm}A=c_4t^3+c_5t^2+c_6t+c_7\;,
	\end{equation}
	and the gauge function $G(x,\dot{x};t)$ as
	\begin{equation}
		\begin{split}
			G=c_13x^2\dot{x}^2+&c_4\left(36x\dot{x}-18\dot{x}^2t^2+4\frac{\dot{x}^3t^3}{x}-18x^2\right)\\
			+&c_5\left(12x\dot{x}-12\dot{x}^2t+4\frac{\dot{x}^3t^2}{x}\right)+c_6\left(-6\dot{x}^2+4\frac{\dot{x}^3t}{x}\right)+c_74\frac{\dot{x}^3}{x}\;.
		\end{split}
	\end{equation}
	Therefore the symmetry generators and corresponding integrals of motion are
	\begin{alignat}{3}
		\label{I1} \Gamma_1&=t^2\frac{\partial}{\partial  t}+\frac{3}{2}xt\frac{\partial}{\partial x}\hspace{.5cm}&&;\hspace{.5cm}I_1&&=3x^3\ddot{x}-x^2\dot{x}^2+(2x\dot{x}^3-7x^2\dot{x}\ddot{x}-3x^3\dddot{x})t \\ \nonumber
		&\text{ }&&\text{ }&&+(-\dot{x}^4-x^2\ddot{x}^2+4x\dot{x}^2\ddot{x}+2x^2\dot{x}\dddot{x})t^2 \\
		\Gamma_2&=t\frac{\partial}{\partial t}+\frac{3}{4}x\frac{\partial}{\partial x}\hspace{.5cm}&&;\hspace{.5cm}I_2&&=2x\dot{x}^3-7x^2\dot{x}\ddot{x}-3x^3\dddot{x}\\ \nonumber
		&\text{ }&&\text{ }&&+(-2\dot{x}^4-2x^2\ddot{x}^2+8x\dot{x}^2\ddot{x}+4x^2\dot{x}\dddot{x})t \\
		\Gamma_3&=\frac{\partial}{\partial t}\hspace{.5cm}&&;\hspace{.5cm}I_3&&=-\dot{x}^4-x^2\ddot{x}^2+4x\dot{x}^2\ddot{x}+2x^2\dot{x}\dddot{x} \\
		\Gamma_4&=\frac{t^3}{x}\frac{\partial}{\partial x}\hspace{.5cm}&&;\hspace{.5cm}I_4&&=3x^2-6x\dot{x}t+(3\dot{x}^2+3x\ddot{x})t^2+(-x\dddot{x}-3\dot{x}\ddot{x})t^3 \\
		\Gamma_5&=\frac{t^2}{x}\frac{\partial}{\partial x}\hspace{.5cm}&&;\hspace{.5cm}I_5&&=-2x\dot{x}+(2\dot{x}^2+2x\ddot{x})t+(-x\dddot{x}-3\dot{x}\ddot{x})t^2 \\
		\Gamma_6&=\frac{t}{x}\frac{\partial}{\partial x}\hspace{.5cm}&&;\hspace{.5cm}I_6&&=\dot{x}^2+x\ddot{x}+(-3\dot{x}\ddot{x}-x\dddot{x})t \\
		\Gamma_7&=\frac{1}{x}\frac{\partial}{\partial x}\hspace{.5cm}&&;\hspace{.5cm}I_7&&=3\dot{x}\ddot{x}+x\dddot{x}\;.
	\end{alignat}
	In order to pass into primed coordinates where $x'$ is a cyclic coordinate, for simplicity we set $c_1 = 1$ and other constants to zero. Then we make a suitable coordinate transformation,
	\begin{equation}
		t =- \frac{1}{x'} \hspace{.5cm};\hspace{.5cm}  x = t'{x'}^{-\frac{3}{2}}\;.
	\end{equation}
	The gauge function $G(x,\dot{x};t)$ in old coordinates is given by the expression
	\begin{equation}
		G= 3x^2 \dot{x}^2\;,
	\end{equation}
	and
	\begin{equation}
		F = 3\frac{{t\prime}^2}{{\dot{x}\prime}^2 {x\prime}} - \frac{9}{2}\frac{{t\prime}^3}{{\dot{x}\prime}{x\prime}^2} + \frac{9}{4}\frac{{t\prime}^4}{{x\prime}^3}\;.
	\end{equation}
	From the equation \eqref{tilde L} the new Lagrangian $\tilde{L}'$ gives to the desired result
	\begin{equation}
		\tilde{L}\prime = 3\frac{{\ddot{x}\prime}^2 {t\prime}^2 }{{\dot{x}\prime}^5} + \frac{1}{{\dot{x}\prime}^3}\;.
	\end{equation}
	Here $x'$ is a cyclic coordinate and the corresponding conserved momentum gives the expression $\eqref{I1}$.

	\subsection{Free particle with a Chern-Simons like term}
	
	Now let us examine a system with two degrees of freedom. We consider the following Lagrangian system introduced by Lukierski, Stichel, and Zakrzewski \cite{1997}
	\begin{equation}\label{chiral}
		L=\frac{\lambda}{2}(\dot{y}\ddot{x}-\dot{x}\ddot{y})+\frac{m}{2}(\dot{x}^2+\dot{y}^2) \;,
	\end{equation}
	where $m$ stands for the mass, and $\lambda$ is some parameter. Since the latter parameter labels the central extension \cite{1997} and has interesting properties, we specially keep the parameters $m$ and $\lambda$. The Euler-Lagrange equations are
	\begin{align}
		m\ddot{x}-\lambda\dddot{y} & =0 \;, \\
		m\ddot{y}+\lambda\dddot{x} & =0 \;.
	\end{align}
	Since in general $\eta_1$, $\eta_2$ and $\zeta$ depend on $x,y$ and $t$ using our prescription one obtains the following three partial differential equations
	\begin{equation}
		\frac{\lambda}{2}\dot{y}\frac{\partial\eta_2}{\partial y}+\frac{\lambda}{2}\frac{\partial\eta_2}{\partial t}+\frac{\lambda}{2}\dot{y}\frac{\partial\eta_1}{\partial x}-\lambda\dot{y}\frac{\mathrm{d}\zeta}{\mathrm{d}t}=\frac{\partial G}{\partial \dot{x}} \;,
	\end{equation}
	\begin{equation}
		-\frac{\lambda}{2}\dot{x}\frac{\partial\eta_1}{\partial x}-\frac{\lambda}{2}\frac{\partial\eta_1}{\partial t}+\lambda\dot{x}\frac{\mathrm{d}\zeta}{\mathrm{d}t}-\frac{\lambda}{2}\dot{x}\frac{\partial\eta_2}{\partial y}=\frac{\partial G}{\partial \dot{y}} \;,
	\end{equation}
	\begin{equation}
		\begin{split}
			m\dot{x}\frac{\mathrm{d}\eta_1}{\mathrm{d}t}+m\dot{y}\frac{\mathrm{d}\eta_2}{\mathrm{d}t}+\frac{\lambda}{2}\dot{y}\Bigr[\dot{x}^2\frac{\partial^2\eta_1}{\partial x^2}+2\dot{x}\dot{y}\frac{\partial^2\eta_1}{\partial x\partial y}+2\dot{x}\frac{\partial^2\eta_1}{\partial x\partial t}+\dot{y}^2\frac{\partial^2\eta_1}{\partial y^2}+2\dot{y}\frac{\partial^2\eta_1}{\partial y\partial t}+\frac{\partial^2\eta_1}{\partial t^2}\Bigr]\\-\frac{\lambda}{2}\dot{x}\Bigr[\dot{x}^2\frac{\partial^2\eta_2}{\partial x^2}+2\dot{x}\dot{y}\frac{\partial^2\eta_2}{\partial x\partial y}+2\dot{x}\frac{\partial^2\eta_2}{\partial x\partial t}+\dot{y}^2\frac{\partial^2\eta_2}{\partial y^2}+2\dot{y}\frac{\partial^2\eta_2}{\partial y\partial t}+\frac{\partial^2\eta_2}{\partial t^2}\Bigr]+\frac{m}{2}(\dot{x}^2+\dot{y}^2)\frac{\mathrm{d}\zeta}{\mathrm{d}t} \\ 
			= \frac{\partial G}{\partial x}\dot{x}+\frac{\partial G}{\partial y}\dot{y}+\frac{\partial G}{\partial t} \;.
		\end{split}
	\end{equation}
	By solving these equations, one finds infinitesimal transformations and the gauge function as
	\begin{equation}
		\zeta=c_1 \;,
	\end{equation}
	\begin{equation}
		\eta_1=c_2y+c_3\sin{(\frac{m}{\lambda}t)}+c_4\cos{(\frac{m}{\lambda}t)}+c_5t+c_6 \;,
	\end{equation}
	\begin{equation}
		\eta_2=-c_2x-c_3\cos{(\frac{m}{\lambda}t)}+c_4\sin{(\frac{m}{\lambda}t)}+c_7t+c_8 \;,
	\end{equation}
	\begin{equation}
		\begin{split}
			G=c_3\left(\frac{m}{2}\dot{x}\sin{(\frac{m}{\lambda}t)}-\frac{m}{2}\dot{y}\cos{(\frac{m}{\lambda}t)}\right)+&c_4\left(\frac{m}{2}\dot{y}\sin{(\frac{m}{\lambda}t)}+\frac{m}{2}\dot{x}\cos{(\frac{m}{\lambda}t)}\right)\\
			+&c_5\left(mx-\frac{\lambda}{2}\dot{y}\right)+c_7\left(my+\frac{\lambda}{2}\dot{x}\right) \;.
		\end{split}
	\end{equation}
	Then the integrals of motion with the corresponding symmetry generators are
	\begin{alignat}{3}
		\Gamma_1&=\frac{\partial}{\partial t}\hspace{.5cm}&&;\hspace{.5cm}I_1&&=\lambda\left(\dot{x}\ddot{y}-\dot{y}\ddot{x}\right)-\frac{m}{2}\left(\dot{x}^2+\dot{y}^2\right) \\ \label{I2}
		\Gamma_2&=y\frac{\partial}{\partial x}-x\frac{\partial}{\partial y}\hspace{.5cm}&&;\hspace{.5cm}I_2&&=\lambda\left(\frac{\dot{x}^2}{2}+\frac{\dot{y}^2}{2}-x\ddot{x}-y\ddot{y}\right)+m\left(\dot{x}y-\dot{y}x\right) \\
		\Gamma_3&=\sin{(\frac{m}{\lambda}t)}\frac{\partial}{\partial x}-\cos{(\frac{m}{\lambda}t)}\frac{\partial}{\partial y}\hspace{.5cm}&&;\hspace{.5cm}I_3&&=\lambda\left(\ddot{y}\sin{(\frac{m}{\lambda}t)}+\ddot{x}\cos{(\frac{m}{\lambda}t)}\right) \\
		\Gamma_4&=\cos{(\frac{m}{\lambda}t)}\frac{\partial}{\partial x}+\sin{(\frac{m}{\lambda}t)}\frac{\partial}{\partial y}\hspace{.5cm}&&;\hspace{.5cm}I_4&&=\lambda\left(\ddot{y}\cos{(\frac{m}{\lambda}t)}-\ddot{x}\sin{(\frac{m}{\lambda}t)}\right) \\
		\Gamma_5&=t\frac{\partial}{\partial x}\hspace{.5cm}&&;\hspace{.5cm}I_5&&=\lambda\left(\dot{y}-\ddot{y}t\right)+m\left(-x+\dot{x}t\right)\\
		\Gamma_6&=\frac{\partial}{\partial x}\hspace{.5cm}&&;\hspace{.5cm}I_6&&=-\lambda\ddot{y}+m\dot{x} \\
		\Gamma_7&=t\frac{\partial}{\partial y}\hspace{.5cm}&&;\hspace{.5cm}I_7&&=\lambda\left(-\dot{x}+\ddot{x}t\right)+m\left(-x+\dot{x}t\right) \\
		\Gamma_8&=\frac{\partial}{\partial y}\hspace{.5cm}&&;\hspace{.5cm}I_8&&=\lambda\ddot{x}+m\dot{y}
	\end{alignat}
	From the form of the Lagrangian \eqref{chiral}, it can be seen that $x$ and $y$ are cyclic coordinates. We will pursue the same method to find the Lagrangian $\tilde{L}'$, where the momentum conjugate corresponding to $x'$ is given by \eqref{Q}. Consider the case in which $c_2=1 $ is the only constant that does not equal to zero. Then
	\begin{equation}
		\frac{\partial t}{\partial x'} = 0 \hspace{.5cm};\hspace{.5cm}   \frac{\partial x}{\partial x'} = y \hspace{.5cm};\hspace{.5cm}   \frac{\partial y}{\partial x'} = -x \hspace{.5cm};\hspace{.5cm} G=0 \;.
	\end{equation}
	It is clear that these symmetries correspond to the transformation to polar coordinates
	\begin{equation}
		t = t' \hspace{.5cm};\hspace{.5cm}  x = y' \sin{x'} \hspace{.5cm};\hspace{.5cm}  y = y' \cos{x'} \; .
	\end{equation}
	After calculating the first and second derivatives of the primed coordinates, we can rewrite our Lagrangian as
	\begin{equation}
		L\prime= \frac{\lambda}{2}\left( {\dot{x}\prime}^3{y\prime}^2 + 2\dot{x}\prime {\dot{y}\prime}^2 +\ddot{x}\prime \dot{y}\prime y\prime - \dot{x}\prime \ddot{y}\prime y\prime \right) + \frac{m}{2}\left({\dot{y}\prime}^2 + {\dot{x}\prime}^2 {y\prime}^2 \right) \;.
	\end{equation}
	Obviously, $x'$ is a cyclic coordinate and the corresponding conserved quantity to the symmetry is as follows
	\begin{equation}
		I_2 = \lambda\left(\frac{\dot{x}^2}{2}+\frac{\dot{y}^2}{2}-x\ddot{x}-y\ddot{y}\right)+m\left(\dot{x}y-\dot{y}x\right)  \;.
	\end{equation}
	which is also given by \eqref{I2}.
	
	\section{Variational Symmetries for High-Order Lagrangians}
	
	Finally, let us generalize what we have done in the previous sections for Lagrangian systems with $N$th order derivatives
	\begin{equation}
		L=L(\mathbf{x},\dot{\mathbf{x}},...,\mathbf{x}^{(N)};t) \; .
	\end{equation}
	Using the same procedure one find the  generalized version of the equation $\eqref{eq1}$ 
	\begin{equation} \label{thirdorder}
		\sum_{i=1}^n\sum_{k=0}^N\frac{\partial L}{\partial x_i^{(k)}}\Bigr[\frac{\mathrm{d}^k}{\mathrm{d}t^k}\Bigr(\eta_i-\dot{x}_i\zeta\Bigr)+x_i^{(k+1)}\zeta\Bigr]+\frac{\partial L}{\partial t}\zeta+L\frac{\mathrm{d}\zeta}{\mathrm{d}t}=\frac{\mathrm{d}}{\mathrm{d}t}G(\mathbf{x},\dot{\mathbf{x}},...,\mathbf{x}^{(N-1)};t) \; .
	\end{equation}
	The generalized form of the equation $\eqref{Q}$ can be written as follows \cite{Woodard:2015zca}
	\begin{equation}
		\frac{\mathrm{d}}{\mathrm{d}t}\left(\zeta L-\sum_{i=1}^n\sum_{k=1}^{N} \sum_{j=1}^{k}(-1)^{j} \frac{\mathrm{d}^{k-j}(\eta_i-\dot{x}_i\zeta)}{\mathrm{d} t^{k-j}} \frac{\mathrm{d}^{j-1}}{\mathrm{d} t^{j-1}}\left(\frac{\partial L}{\partial x_i^{(k)}} \right)-G\right)=0 \;.
	\end{equation}
	As an example let us consider the system with the following Lagrangian
	\begin{equation}
		L=\frac{\dddot{x}^2}{2} \; . 
	\end{equation}
	The Euler-Lagrange equation of the system is the sixth order ordinary differential equation
	\begin{equation}
		x^{(6)}=0 \; . 
	\end{equation}
	Substituting the Lagrangian into the expression (\ref{thirdorder}) we obtain the following result
	\begin{equation}
		\zeta(t)=c_1t^2+c_2t+c_3 
	\end{equation}
	\begin{equation}
		\eta(x,t)=c_15xt+c_2\frac{5}{2}x+c_4t^5+c_5t^4+c_6t^3+c_7t^2+c_8t+c_9 
	\end{equation}
	\begin{equation}
		G(x,\dot{x},\ddot{x},t)=c_1\frac{9}{2}\ddot{x}^2+c_4\left(60\ddot{x}t^2-120\dot{x}t+120x\right)+c_5\left(24\ddot{x}t-24\dot{x}\right)+c_66\ddot{x} \; . 
	\end{equation}
	Using these expressions we find symmetry generators and integrals of motion 
	\begin{alignat}{3}
		\Gamma_1 &= t^2\frac{\partial}{\partial t}+5xt\frac{\partial}{\partial x}\hspace{.5cm}&&;\hspace{.5cm}I_1&&=-\frac{9}{2}\ddot{x}^2+8\dot{x}\dddot{x}-5xx^{(4)} \\ &\text{ }&&\text{}&&+(\ddot{x}\dddot{x}-3\dot{x}x^{(4)}+5xx^{(5)})t+(-\frac{\dddot{x}^2}{2}+\ddot{x}x^{(4)}-\dot{x}x^{(5)})t^2 \\
		\Gamma_2 &=t\frac{\partial}{\partial t}+\frac{5}{2}x\frac{\partial}{\partial x}\hspace{.5cm}&&;\hspace{.5cm}I_2&&=\ddot{x}\dddot{x}-3\dot{x}x^{(4)}+5xx^{(5)}+(2\ddot{x}x^{(4)}-2\dot{x}x^{(5)}-\dddot{x}^2)t \\
		\Gamma_3&=\frac{\partial}{\partial t}\hspace{.5cm}&&;\hspace{.5cm}I_3&&=-\dddot{x}^2+2\ddot{x}x^{(4)}-2\dot{x}x^{(5)} \\
		\Gamma_4&=t^5\frac{\partial}{\partial x}\hspace{.5cm}&&;\hspace{.5cm}I_4&&=-120x+120\dot{x}t-60\ddot{x}t^2+20\dddot{x}t^2-5x^{(4)}t^4+x^{(5)}t^5 \\
		\Gamma_5&=t^4\frac{\partial}{\partial x}\hspace{.5cm}&&;\hspace{.5cm}I_5&&=24\dot{x}-24\ddot{x}t+12\dddot{x}t^2-4x^{(4)}t^3+x^{(5)}t^4 \\
		\Gamma_6&=t^3\frac{\partial}{\partial x}\hspace{.5cm}&&;\hspace{.5cm}I_6&&=-6\ddot{x}+6\dddot{x}t-3x^{(4)}t^2+x^{(5)}t^3 \\
		\Gamma_7&=t^2\frac{\partial}{\partial x}\hspace{.5cm}&&;\hspace{.5cm}I_7&&=2\dddot{x}-2x^{(4)}t+x^{(5)}t^2 \\
		\Gamma_8&=t\frac{\partial}{\partial x}\hspace{.5cm}&&;\hspace{.5cm}I_8&&=-x^{(4)}+x^{(5)}t \\
		\Gamma_9&=\frac{\partial}{\partial x}\hspace{.5cm}&&;\hspace{.5cm}I_9&&=x^{(5)}  \; . 
	\end{alignat}

	\section{Conclusions}
	
 This paper dealt with an elementary derivation of integrals of motion for higher-order Lagrangian systems based on variational symmetries technique. The methods used here to compute variational symmetries and related integrals of motion are an extension of those used in \cite{torres2013variational} and \cite{del2017variational}. We provided variational symmetries of higher-order Lagrangian systems which are well-known in the literature and studied from different viewpoints.

The symmetries discussed here can be useful also in the context of the quantization of the higher derivative Lagrangian systems. Another possible direction is the derivation of symmetries for field theories with higher-derivative terms.

\section*{Acknowledgements} 
		
		We would like to thank all the participants of the seminar series on ``Higher-derivative systems'' held at Bogazici University in summer 2020. The work of Ilmar Gahramanov is partially supported by the Bogazici University Research Fund under grant number 20B03SUP3. Ege \c{C}oban and Dilara Kosva are supported by the 2209-A TUBITAK National/International Research Projects Fellowship Programme for Undergraduate Students under grant number 1919B012000987.

%
%
%
\bibliographystyle{utphys}
	\bibliography{varsym}

\end{document}